\documentclass[aps,prl,twocolumn]{revtex4}

\usepackage{graphicx}
\usepackage{dcolumn}
\usepackage{bm}

\begin{document}
\title{Joule heating induced negative differential resistance in free standing metallic carbon nanotubes}


\author{Marcelo A. Kuroda}
\affiliation{Beckman Institute and Department of Physics,University of Illinois Urbana-Champaign, IL 61801}
\author{Jean-Pierre Leburton}
\email{jleburto@uiuc.edu}
\affiliation{Beckman Institute and Department
of Electrical and Computer Engineering, University of Illinois Urbana-Champaign, IL 61801}
\date{\today}

\begin{abstract}
The features of the $IV$ characteristics of metallic carbon
nanotubes (m-NTs) in different experimental setups are studied
using semi-classical Boltzmann transport equation together with
the heat dissipation equation to account for significant thermal
effects at high electric bias. Our model predicts that the shape
of the m-NT characteristics is basically controlled by heat
removal mechanisms. In particular we show that the onset of
negative differential resistance in free standing nanotubes finds
its origins in strong transport nonlinearities associated with
poor heat removal unlike in substrate-supported nanotubes.
\end{abstract}

\maketitle Since their discovery~\cite{ijima}, carbon nanotubes (NTs)
have captured the attention of both the scientific and
technological communities because of their mechanical stability as
well as their high thermal and electrical
capabilities~\cite{saito_dres} not usually seen in other
materials. Among their outstanding properties is the peculiarity
to behave as metals or semiconductors depending on their chirality
\cite{saito,mintmire}. Their capability of carrying large current
densities at room temperature make them prominent materials for
field effect transistors \cite{tans,martel} and interconnects
\cite{mceuen} in high speed nanoscale electronics. The first
measurements of the $IV$ characteristics on individual metallic
nanotubes (m-NTs) \cite{yao} were performed on substrate-supported
tubes in a configuration similar to a field effect transistor. The
output characteristics resembled those obtained in semiconductors
as they exhibited linear dependence on voltage in the low field
regime, and saturation at $I\!\sim\! 25\mu\mbox{A}$ under high
biases. Later measurements have shown that in short m-NTs, the
saturation level could be overcome, but a nonlinear behavior
still persists in substrate-supported configurations
\cite{javey,park}. In recent experiments performed on m-NTs
standing freely across a trench, current levels in the high bias
regime were a few times smaller than those in nanotubes supported
by substrates while negative differential resistance (NDR) was
reported \cite{pop}. This behavior, indicating significant heat
production/dissipation in m-NTs devices, was interpreted in terms
of the onset of a non-equilibrium optical phonon population.

\begin{figure}[htbp]
  \leavevmode \centering
    \includegraphics[width=3.in]{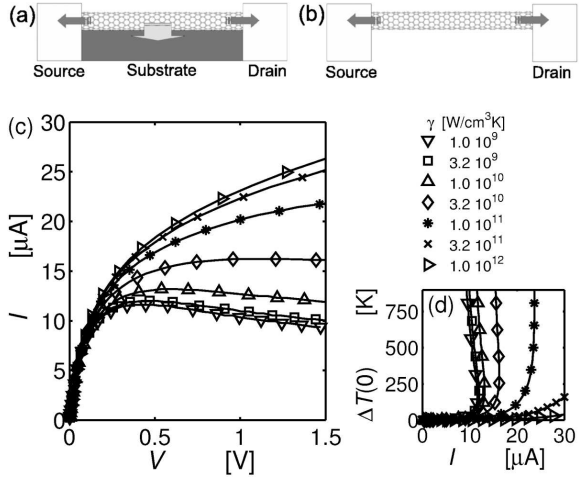}
    \vspace{-.15in}
  \caption{\label{devices} Experimental setups: (a) Substrate-supported
  nanotube (SST); (b) Free-standing nanotube (FST). (c) $IV$
  characteristics for 800nm-long nanotube with different
  values of coupling coefficient $\gamma$. (d) Temperature in the
  middle of the tube $\Delta T(0)$ as a function of the current $I$ for different
  $\gamma$.}
\end{figure}

In this Letter we show that the nonlinearities in the $IV$
characteristics of m-NTs, find their origins in Joule heating and
the efficiency in heat removal from the m-NT. We determine the
temperature profile along nanotubes by solving the Boltzmann
transport equation simultaneously with the heat transfer equation.
We specifically demonstrate that while the nonlinear behavior in
substrate-supported tubes (SSTs) (Fig.~\ref{devices}.a) emerges
due to a strong imbalance between carrier distribution functions
with positive and negative Fermi velocity, the characteristics in
free-standing tubes (FSTs) (Fig.~\ref{devices}.b) essentially
arise from an inhomogeneous self-heating effect.

The distance between sub-bands in m-NTs using tight-binding
calculations can be estimated as $\Delta E_{met} \!\approx\! 6
\gamma_0a/D$ where $a\!=\!0.14nm$, $\gamma_0 \!\approx\!3$eV and
$D$ is the diameter of the tube \cite{venema}. Hence, for
transport purposes, the electronic structure of small diameter
m-NTs close to the Fermi level can be well described by linear
dispersion relations $E(k) = \pm \hbar v_F (k-k_F)$ with a Fermi
velocity $v_F$ close to $8.10^7$cm/s. This approximation is valid
as long as temperature satisfies that $k_BT \!\ll \!\Delta
E_{met}$ and the bias voltage $V\! <\! \Delta E_{met}$. The
current along the nanotube is naturally defined as:
\begin{equation}
I = e\, v_F (n^+-n^-) \label{currentdef}
\end{equation}
where $n^+$ ($n^-$) is the electron density with positive
(negative) Fermi velocity. Similarly to Ref.~\cite{kuroda}, we
assume that carrier populations are described by Fermi statistics
with different quasi-Fermi level $E_F^\alpha$ depending on the sign $\alpha$ of the
Fermi velocity. Using the zeroth moment of the Boltzmann equation,
the electric field $F$ satisfies:
\begin{equation}
v_F \partial_x\! \left(n^+\!\! +\! n^-\!\right)\!
-\!\frac{2eF}{\pi \hbar} = 2 \tilde{C}_{ph}\!(I,T).
\label{boltzmann2}
\end{equation}
Here $\tilde{C}_{ph}\!(I,T)$ is the momentum integral of the
collision integral considering only the contribution to scattering
involving high energy optical phonons ($\hbar\omega\sim0.18$eV).
In our model, phonons are assumed to be in {\it local} thermal
equilibrium with electrons, and their occupation number follows
the Bose-Einstein statistics.

A realistic description of the $IV$ characteristics requires the
consideration of the temperature profile $T(x)$ along the m-NT in
the right hand side of Eq.~\ref{boltzmann2}. We account for this
local temperature variation by solving the heat
production/dissipation equation:
\begin{equation}
-\frac{d}{dx}\left(\kappa \frac{dT}{dx}\right) + \gamma
\left(T-T_0\right) = q^*, \label{heateq}
\end{equation}
in which $\kappa$, $\gamma$, $T_0$, and $q^*$ are the thermal
conductivity, the thermal coupling to the substrate, the substrate
temperature, and the heat dissipation per unit volume,
respectively.  The first term on the left hand side of
Eq.~\ref{heateq} corresponds to the heat flowing through the
leads; the second, to the flow driven through the substrate or any
surrounding medium. The heat produced locally is assumed to obey
Joule's law $q^*\!=\!I F/\pi D t$, being $t$ the effective
thickness of the tube ($t\approx 0.34$nm). The m-NT thermal
conductivity dependence on temperature has been predicted to
follow $\kappa(T)\! =\! \kappa_{RT} T_{RT}/T$, due to the Umklapp
process of acoustic phonons~\cite{osman} for $T\! \gtrsim\! 300K$.
In this equation, $\kappa_{RT}$ is the thermal conductivity at a
reference temperature $T_{RT}$ ($\kappa\! \sim\! 20\mbox{W/cmK}$
at 300K). An estimate of the thermal coupling constant $\gamma$ of
nanotubes standing on SiO$_2$ substrates is
$10^{11}\mbox{W/cm}^3\mbox{K}$ \cite{kuroda}. The explicit
dependence of the temperature on the local field requires
Eqs.~\ref{boltzmann2} and \ref{heateq} to be solved
simultaneously, leading to a non-linear second order differential
equation for the temperature. We set the boundary conditions as
$T(\pm L/2)\!=\!T_0$, where $L$ is the m-NT length, assuming that
both contact leads have the same temperature as the substrate
(i.e. we neglect the heating at the leads due to a contact
resistance).

In Fig.~\ref{devices}(c) we compare the performances of an 800nm
long m-NT assuming different $\gamma$ values. The parameters used
in the calculations are $\kappa\!=\!20$W/cmK, $\tau = 22$fs and
$\hbar \omega\!=\!0.18$eV. For the smallest $\gamma$ values, the
m-NT thermal coupling to the substrate is vanishingly small (less
than 10\% of the power generated along the tube is dissipated in
the high bias through the substrate) and NDR is observed in the
$IV$ characteristics. As we increase $\gamma$ to $\sim\!
10^{11}\mbox{W/cm}^3\mbox{K}$ saturation is observed. If we
further increase $\gamma$, the saturation is overcome but
deviations from the linear regime still arise. Our results clearly
indicate that in the low bias regime, thermal effects can be
neglected, but the high bias regime is basically governed by the
m-NT capability for heat removal. Hence, the features of the
electrical characteristics depend not only on the tube's length
but also on $\gamma$. In Fig.~\ref{devices}(d) we plot the
temperature difference $\Delta T(0)\!=\!T(0)-T_0$  at the m-NT
midlength as a function of the current for the set same of
$\gamma$-values. Our model predicts that the larger the coupling
coefficient $\gamma$, the higher current in the m-NT.

\begin{figure}[htbp]
  \leavevmode \centering
    \includegraphics[width=3.in]{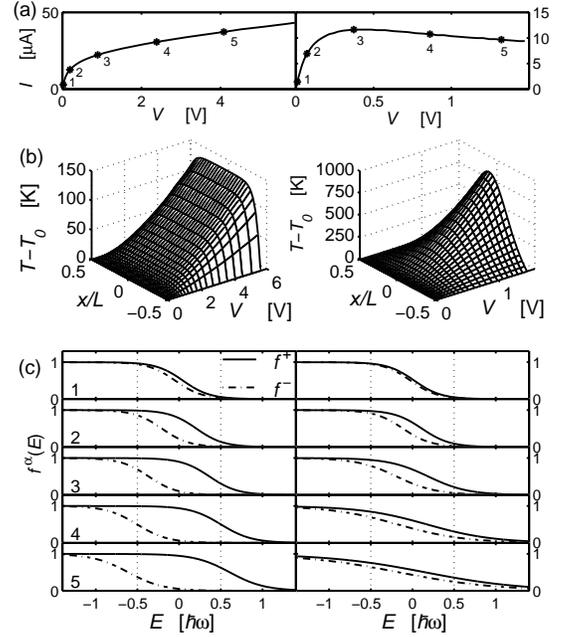}
  \caption{\label{iv_coup} Features comparison of 800nm long
    m-NT strongly (left) and weakly (right) coupled to the
   substrate: (a) $IV$ characteristics;
  (b) Temperature profile $\Delta T(x)$ as a function of the voltage $V$;
  (c) Fermi distribution at midlength for each of the carrier
branches at different points indicated in the $IV$
characteristics. The distance between dashed line corresponds to
the optical phonon energy.}
\end{figure}

In Fig.~\ref{iv_coup}, we consider the situation of an 800nm long
m-NT in two thermal coupling regimes i.e. i) strongly coupled (SST
with $\gamma= 10^{12}\mbox{W/cm}^3\mbox{K}$, left hand side of
Fig.~\ref{iv_coup}) and ii) weakly coupled (FST with $\gamma =
10^9\mbox{W/cm}^3\mbox{K}$, right hand side of Fig.~\ref{iv_coup})
to the substrate. In the SST, the $IV$ characteristic exhibits a
nonlinear behavior with high resistance at high fields
(Fig.~\ref{iv_coup}.a. LHS). In this case, the substrate acts as a
heat sink, providing efficient dissipation of the heat produced
along the tube while keeping both the electric field and power
dissipated per unit volume low. This efficient heat removal
reduces the temperature rise in the m-NT, thereby favoring
relatively high current levels while maintaining a quasi uniform
temperature profile along the tube (except at the edges) as shown
in the LHS of Fig.~\ref{iv_coup}.b. The LHS of
Fig.~\ref{iv_coup}.c displays the profile of the (Fermi-like)
carrier distributions at the NT mid-length for electrons with
positive and negative Fermi velocity ($f^+$  and $f^-$) at the
different points indicated along the $IV$ curve in
Fig.~\ref{iv_coup}.a (LHS). No significant heating is observed
before the separation between the two distribution quasi-Fermi
levels becomes comparable to the optical phonon energy
($E_F^+\!-\!E_F^-\!\approx\!\hbar\omega$). Hence, the nonlinear
behavior observed in this kind of configuration is basically
attributed to the smooth onset of optical phonon scattering \cite
{yao,kuroda}, despite minor thermal broadening induced by Joule
heating on the carrier distributions. In the FST, the $IV$
characteristic exhibits a NDR at relatively low biases
(Fig.~\ref{iv_coup}.a RHS), and the current level is lower than in
the previous case of strong thermal coupling. Because of poor heat
removal, temperature rises rapidly with external biases
manifesting a pronounced maximum at NT mid-length (RHS of
Fig.~\ref{iv_coup}.b). One notices the different scales on the
voltage and temperature axis between the SST (LHS) and FST (RHS).
Simultaneously, the Fermi distributions $f^+$ and $f^-$ remain
weakly separated but experiencing significant heating even at low
voltages as shown in Fig.~\ref{iv_coup}.c (RHS) for the set of
points along the $IV$ characteristic on Fig.~\ref{iv_coup}.a
(RHS). The fast rise of temperature, consequence of the poor heat
removal, locally broadens the electron distributions, and
therefore enhances the scattering locally even at low bias. The
rapid onset of non-uniform $T$-profile induces nonlinearities in
the electric field, which result in a current diminution in order
to satisfy the boundary conditions on the applied voltage at the
NT contacts.

\begin{figure}[htbp]
  \leavevmode \centering
    \includegraphics[width=3.in]{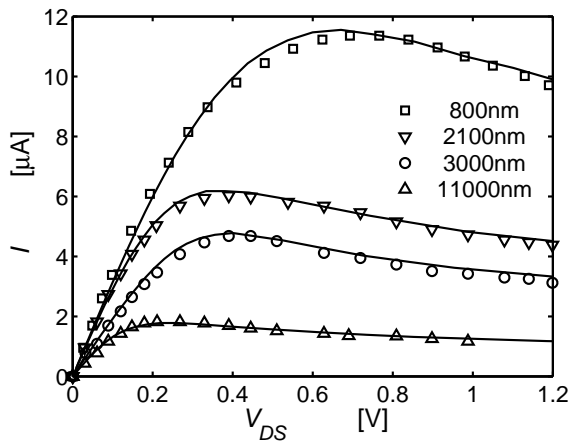}
    \vspace{-.15in}
  \caption{\label{ivchar} Comparison between the $IV$ characteristics for FSTs
of different length calculated with our model (lines) and
experimental data (symbols) of Ref.~\cite{pop}.}
\end{figure}

In Fig.~\ref{ivchar} we compare the results of our model for FSTs
with the experimental data of Ref.~\cite{pop}, for a which good
agreement is observed for all NT-lengths. The physical parameters
used in the calculations are $\hbar \omega\! = \!0.18$eV and
$\kappa_{RT}\! =\! 20$W/cmK. The relaxation time ($\tau\!\sim\!
30$fs) used to compute the characteristics increases with diameter
consistently with theoretical predictions \cite{lazzeri}, and lies
within previous experimental estimates \cite{yao,javey,park}. Our
model also assumes the presence of a contact resistance ($R_c \sim
20\mbox{k}\Omega$).

In conclusion our model shows that the high field transport
properties of m-NTs are strongly controlled by the onset of
thermal effects, which can be altered by modifying the
experimental setup. Using realistic parameters, our model is able
to reproduce the $IV$ characteristics in both substrate-supported
and free-standing nanotubes. However the nonlinear behavior
observed in each of the systems is caused by different phenomena
affecting the carrier distribution. In the former, the transport
features are due to an imbalance in the carrier population with
positive and negative Fermi velocities associated with strong
scattering and high current levels. In the latter, the NDR is
attributed to strong scattering enhancement due to the broadening
of the carrier distribution with low current level.

This work was supported by the Beckman Institute and Network for
Computational Nanotechnology under NSF Grant \# ECC-0228390.

\end{document}